\documentclass[a4paper,11pt]{article}    
\pdfoutput=1
\usepackage[OT1]{fontenc} 
\usepackage{multicol}
\usepackage{graphicx}  
\usepackage{bbold} 
\usepackage{mathtools}
\usepackage{amsmath}
\numberwithin{equation}{section} 
\usepackage{amsfonts} 
\usepackage{amssymb}
\usepackage{amsbsy}
\usepackage{amsfonts}
\usepackage{physics}
\usepackage{bbold}
\usepackage{slashed}  
\usepackage{color}
\usepackage[table]{xcolor}
\usepackage{tablestyles} 
\usepackage{tabularx}
\usepackage{hyperref}
\usepackage{orcidlink}
\usepackage{bbding}

\definecolor{oucrimsonred}{rgb}{0.6, 0.0, 0.0}
\definecolor{DarkGray}{gray}{0.4}
\definecolor{forestgreen}{rgb}{0.13,0.35,0.13}
\definecolor{ocre}{HTML}{F16723}

\usepackage[page,toc,titletoc,title]{appendix}

\numberwithin{equation}{section}
\numberwithin{table}{section}
\numberwithin{figure}{section}
\usepackage{bm}
\usepackage{cite}  
\usepackage{mathrsfs}
\usepackage{framed}
\usepackage[font=footnotesize,labelfont=bf]{caption}
\def\eq#1{{Eq.~(\ref{#1})}}
\def\eqs#1#2{{Eqs.~(\ref{#1})--(\ref{#2})}}

\def\abs#1{\left| #1\right|}

\def\Tr{\mbox{Tr}\,}

\newcommand{\di}{\mbox{d}}

\def\di{\mbox{d}}

\colorlet{grayline}{gray!70}
\definecolor{blueline}{rgb}{0,0.27,0.55}
\definecolor{DarkGray}{gray}{0.4}
\definecolor{Gray}{gray}{0.6}
\definecolor{oucrimsonred}{rgb}{0.6, 0.0, 0.0}
\definecolor{persianblue}{rgb}{0.11, 0.22, 0.73}
\definecolor{forestgreen}{rgb}{0.13,0.35,0.13}
 \hypersetup{colorlinks, citecolor=forestgreen, linkcolor=forestgreen, urlcolor=forestgreen}
\newcommand{\be}{\begin{equation}}
\newcommand{\ee}{\end{equation}}
\newcommand{\bea}{\begin{eqnarray}}
\newcommand{\eea}{\end{eqnarray}}

\newcommand{\CC}{\operatorname{C}}
\newcommand{\BB}{\operatorname{B}}
\newcommand{\hk}{{\bf k}}
\newcommand{\hr}{{\bf r}}
\newcommand{\hn}{{\bf n}}
\newcommand{\hp}{{\bf p}}

\newcommand*\xbar[1]{%
  \hbox{\;%
    \vbox{%
      \hrule height 0.5pt 
      \kern0.5ex
      \hbox{%
        \kern-0.25em
        \ensuremath{#1}%
        \kern-0.07em
      }%
    }%
  }%
} 
\newcommand{\com}[1]{}
\newcommand{\gsim}{\lower.7ex\hbox{$\;\stackrel{\textstyle>}{\sim}\;$}}
\newcommand{\lsim}{\lower.7ex\hbox{$\;\stackrel{\textstyle<}{\sim}\;$}} 

\newcommand{\bc}{\begin{center}}
\newcommand{\ec}{\end{center}}

\newcommand{\blue}[1]{\textcolor{black}{#1}}

\font\beeg=cmr17 scaled 1800
\newbox\ibox
\def\versal#1{\setbox\ibox=\hbox{{\beeg #1}~}%
	    \noindent\global\hangindent=\wd\ibox\global\hangafter-2%
	    \sc\smash{\llap {\lower 14pt \box\ibox}}}
\begin{document}

\hypersetup{citecolor = forestgreen,
linktoc = section, 
linkcolor = forestgreen, 
urlcolor = forestgreen
}

\thispagestyle{empty}
\begin{center}
{ \Large \color{oucrimsonred} \textbf{ 
Quantum tomography with $\tau$ leptons at the FCC-ee\\[+0.3cm]
{\small Entanglement, Bell inequality violation, $\sin \theta_W$ and anomalous couplings}}}

\vspace*{1.5cm}
{\color{black}
  {\bf M. Fabbrichesi$^{a} \orcidlink{0000-0003-1937-3854}$, and} 
 {\bf L. Marzola$^{b,c}\orcidlink{0000-0003-2045-1100}$}
}\\

\vspace{0.5cm}
{\small 
{\it  \color{DarkGray} $^{a}$INFN, Sezione di Trieste, Via Valerio 2, I-34127 Trieste, Italy}
  \\[1mm]  
  {\it \color{DarkGray}$^{b}$Laboratory of High-Energy and Computational Physics, NICPB, R\"avala 10,  10143 Tallinn, Estonia}
  \\[1mm]  
  {\it \color{DarkGray}$^{c}$Institute of Computer Science, University of Tartu, Narva mnt 18, 51009 Tartu, Estonia
  }
  }
\ec 

 \vskip0.5cm
\bc
{\color{DarkGray}
\rule{0.7\textwidth}{0.5pt}}
\ec
\vskip1cm
\bc
{\bf ABSTRACT} 
\ec
The Future Circular Collider (FCC)---in its first incarnation as a lepton collider---will produce, according to the proposed design,  more than 100 billion pairs of $\tau$ leptons after working for four years at the energy of the $Z$-boson resonance. The $\tau$ lepton is special because its relatively long lifetime  makes it possible to reconstruct the momenta of neutrinos emitted in the single pion decay mode. The resulting large number of events is an ideal source for a full quantum tomography of the process that will test quantum entanglement and the violation of Bell inequality with unprecedented precision. In addition, the study of polarizations and spin correlations can provide a competitive determination of the  Weinberg angle $ \theta_W$ and constrain possible anomalous couplings in the neutral electroweak current. We utilize analytic results and  Monte Carlo simulations to explore to what extent these goals might be accomplished.

\vspace*{5mm}

\noindent

  \vskip 3cm
\bc 
{\color{DarkGray} 
\SquareShadowBottomRight
}
\ec

	\newpage

	\tableofcontents
	
	

\newpage
\section{Motivations\label{sec:intro}} 

{\versal The study of the  $\tau$ lepton} is interesting for at least three reasons. For a start, it is the heaviest among the leptons and for this reason alone a particle that deserves a detailed study. Next, an exact reconstruction of the neutrino momentum---and thereby that of the $\tau$ itself--- is made possible by the relatively long lifetime that yields a measurable impact parameter for the charged pions produced in the semi-leptonic decay. Moreover, it decays into hadrons whose angular distribution of momenta allows for the reconstruction of the $\tau$ leptons polarization. The last two of these features make  the quantum tomography of the corresponding polarization density matrix not only possible but also very precise.

The FCC-ee~\cite{FCC:2018evy} is expected to produce $3\times 10^{12}$ visible $Z$ after working for 4 years a the $Z$-boson resonance---of these $1.3 \times 10^{11}$ will decay into $\tau$-lepton pairs. Given the branching ratio of $10.82$\% for the single pronged charged pion plus a neutrino decay mode of the $\tau$, we expect $1.7 \times 10^{9}$ events originated by the produced $\tau$ pairs. These events are the simplest to analyze, have very little if any background and  also provide optimal resolution to analyze the polarization of the decaying leptons. For comparison, and to have a sense of the possible accomplishment, SuperKEKB~\cite{Akai:2018mbz} will produce by the end of its present run $5 \times 10^{10}$ $\tau$ lepton pairs~\cite{Belle-II:2018jsg}, while LEP produced about 137 000 of them~\cite{L3:1998oan}. For these reasons, the FCC-ee is the ideal laboratory for a detailed study of the $\tau$ lepton and of its properties fully accessible via quantum tomography.

The production of the $\tau$ pairs at the $Z$-boson resonance unveils the physics of the electroweak interactions and, in particular, the parity violating interactions yielding non-vanishing polarizations in the final state. These polarizations vanish in the lower energy regime where the photon dominates the production process.  The interplay between polarizations and the other entries in the density matrix shows the origin, and the limited validity, of the maximum entanglement principle featured in purely electromagnetic processes---when only leptons are present. It also makes it possible to determine the Weinberg angle with great precision.

Finally, the definition of appropriated operators in terms of the polarization density matrix, leads to strong constraints on the size of the form factors of all operators up to dimension five affecting the electroweak neutral current. In the low-energy limit, these form factors correspond to the effective radius and magnetic and electric dipole momenta of the $\tau$ lepton. 

Together with the above, perhaps more mundane goals, the FCC could push the study of quantum mechanics to regimes where it has yet to be fully investigated, namely at very high energies  and in the presence of strong and electroweak interactions, opening the possibility of unexpected new fundamental discoveries.

\section{Methods}

{\versal The density matrix of a quantum state} can be fully reconstructed via a procedure dubbed quantum tomography. The $\tau$ leptons---whose spin states are represented by \textit{qubits}, that is, two-level quantum states---act as their own polarimeters and the full polarization density matrix of a $\tau$-lepton pair can be reconstructed by studying the angular distribution of suitable decay products. The simplest and most powerful---in terms of polarization resolution--- of these decay modes is the single-prong $\tau \to \pi \nu_\tau$ decay. Other decay channels, including those yielding three pions or $\rho$ or $a_{1}$ mesons, can also be used for quantum tomography purposes.

\subsection{Quantum tomography at work}

The density matrix describing the polarization state of a quantum system composed by two fermions can be written as
	\begin{equation}
	\label{eq:rho}
		\rho = \frac{1}{4} \qty[
		\mathbb{1}\otimes\mathbb{1} 
		+ 
		\sum_i \BB_i^+ \, \qty(\sigma_i \otimes \mathbb{1}) 
		+ 
		\sum_j \BB_j^- \, \qty(\mathbb{1} \otimes \sigma_j )
		+
		\sum_{i,j} \CC_{ij} \, \qty(\sigma_i \otimes \sigma_j)
		],
	\end{equation}
with $i,j=r, \,n, \,k$ and $\sigma_i$ being the Pauli matrices. The decomposition refers to a right-handed orthonormal basis, $\{\hn, \hr, \hk\}$ and the quantization axis for the polarization is taken along $\hk$, so that $\sigma_k\equiv\sigma_3$. In the fermion-pair center of mass frame we have
	\begin{equation}
		\hn = \frac{1}{\sin \Theta }\qty(\hp \times\hk), \quad \hr = \frac{1}{\sin \Theta }\qty(\hp-  \hk \cos \Theta)\,,
	\end{equation}
where $\hk$ is the direction of the $\tau^+$ momentum and $\Theta$ is the scattering angle. We take $\hp\cdot\hk = \cos\Theta$, with $\hp$ being the direction of the incoming $e^+$. 

The coefficients $\BB^\pm_i$ in \eq{eq:rho} give the polarizations of the individual fermions, whereas the coefficients $\CC_{ij}$ encode the spin correlations. By using $\Tr(\sigma_i\sigma_j)=2\delta_{ij}$ and $\Tr(\sigma_i)=0$, we have:
	\be
		\BB_i^+ = \Tr[\rho\qty(\sigma_i \otimes \mathbb{1}) ]\,,\quad
		\BB_i^- = \Tr[\rho\qty( \mathbb{1} \otimes \sigma_i) ]\,,\quad
		\CC_{ij} = \Tr[\rho\qty(\sigma_i \otimes \sigma_j)]\,.
	\ee
More details on these definitions and quantum tomography in general can be found in~\cite{Barr:2024djo}.

The quantum tomography of the polarization density matrix is completed once the coefficients $\BB_i$ and $\CC_{ij}$ have been found. Given a Lagrangian for a theory, these quantities can be computed from the scattering amplitudes describing the underlying process. Experimentally, or in Monte Carlo simulations, they can instead be reconstructed by tracking the angular distribution of suitable $\tau$-pair decay products. In particular, for events where each $\tau$ lepton decays to a single pion and a neutrino, we have
\begin{equation}
  \frac{1}{\sigma} \,\frac{\dd\sigma}{\dd\cos\theta^\pm_i} = \frac{1}{2}\,
  \qty(1\mp\BB^\pm_i\cos\theta^\pm_i) \,,
\end{equation}

\begin{equation}
  \frac{1}{\sigma} \, \frac{\dd\sigma}{\dd\cos\theta^+_{i} \, \dd\cos\theta^-_{j}} = \frac{1}{4} \, \left( 1 + \CC_{ij} \, \cos\theta^+_{i} \, \cos\theta^-_{j} \right) \, ,
  \label{eq:xsec2d}
\end{equation}
in which $\cos\theta^\pm_{i}$ are the projections of the $\pi^\pm$ momentum direction on the $\{\hn, \hr, \hk\}$ basis, as computed in the rest frame of the decaying $\tau^\pm$. Figures~\ref{fig:Bi} and \ref{fig:Cij} show the 
relative frequencies of the values obtained for the \blue{involved cosines and cosine products from the actual Monte Carlo simulation we have used in this work. All the histograms are normalized to the total number, $10^7$, of simulated  events. The average values of these these quantities, proportional to the $\BB_i$ and $\CC_{ij}$ coefficients, encode the tomographic information and do not vanish provided the corresponding histograms are not symmetric with respect to the zero value.} 

\begin{figure}[ht!]
\begin{center}
\includegraphics[width=5in]{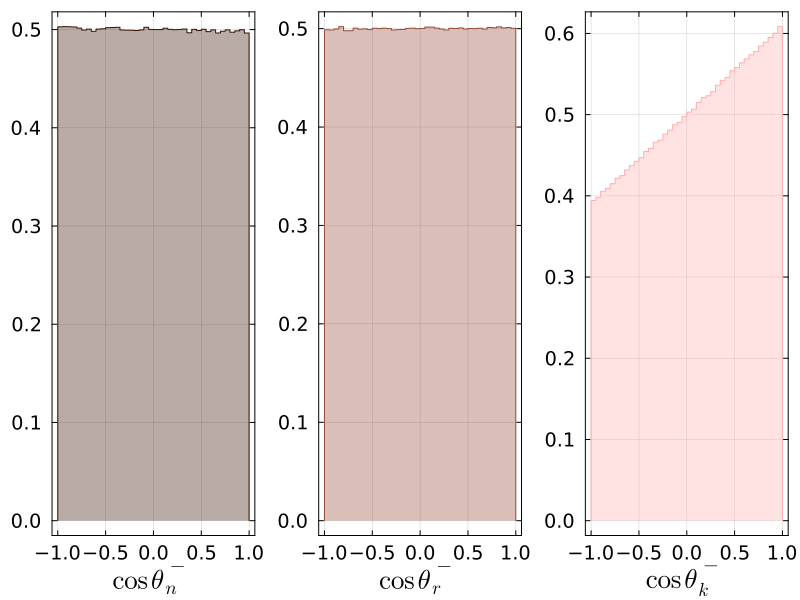}
\vskip 0.2cm
\includegraphics[width=5in]{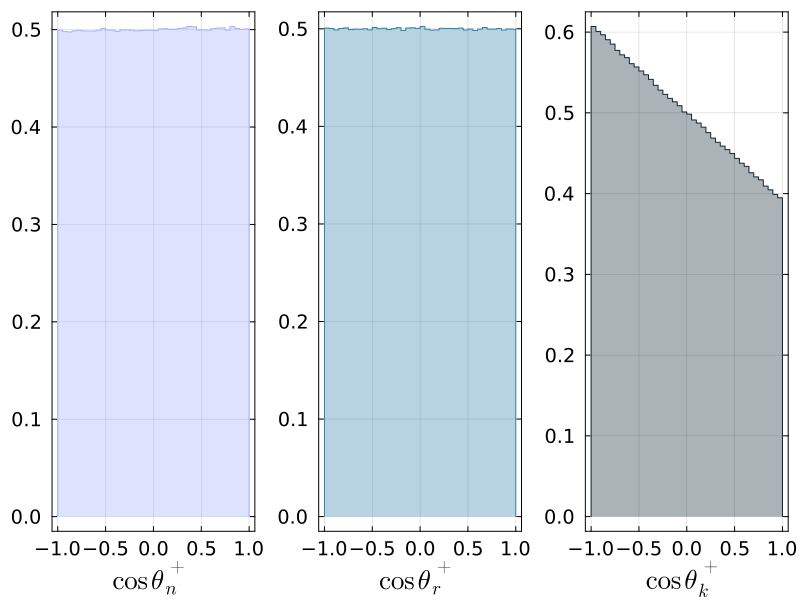}
\caption{\footnotesize Relative frequencies of the values obtained for \blue{ $\cos\theta_i^\pm$, $i=n,r,k$,} in the Monte Carlo simulation of the $\tau$-lepton pairs decaying into a charged pion plus a neutrino, the momentum of which is reconstructed. The histograms are normalized to the total number of events.
\label{fig:Bi} 
}
\end{center}
\end{figure}

Whenever the average of the product of two cosine (in the $\CC_{ij}$ coefficients)  differ from the product of the averages of the single cosines (in the $\BB_i$ coefficients) we have non vanishing correlation and, possibly, quantum entanglement.

\begin{figure}[ht!]
\begin{center}
\includegraphics[width=6.5in]{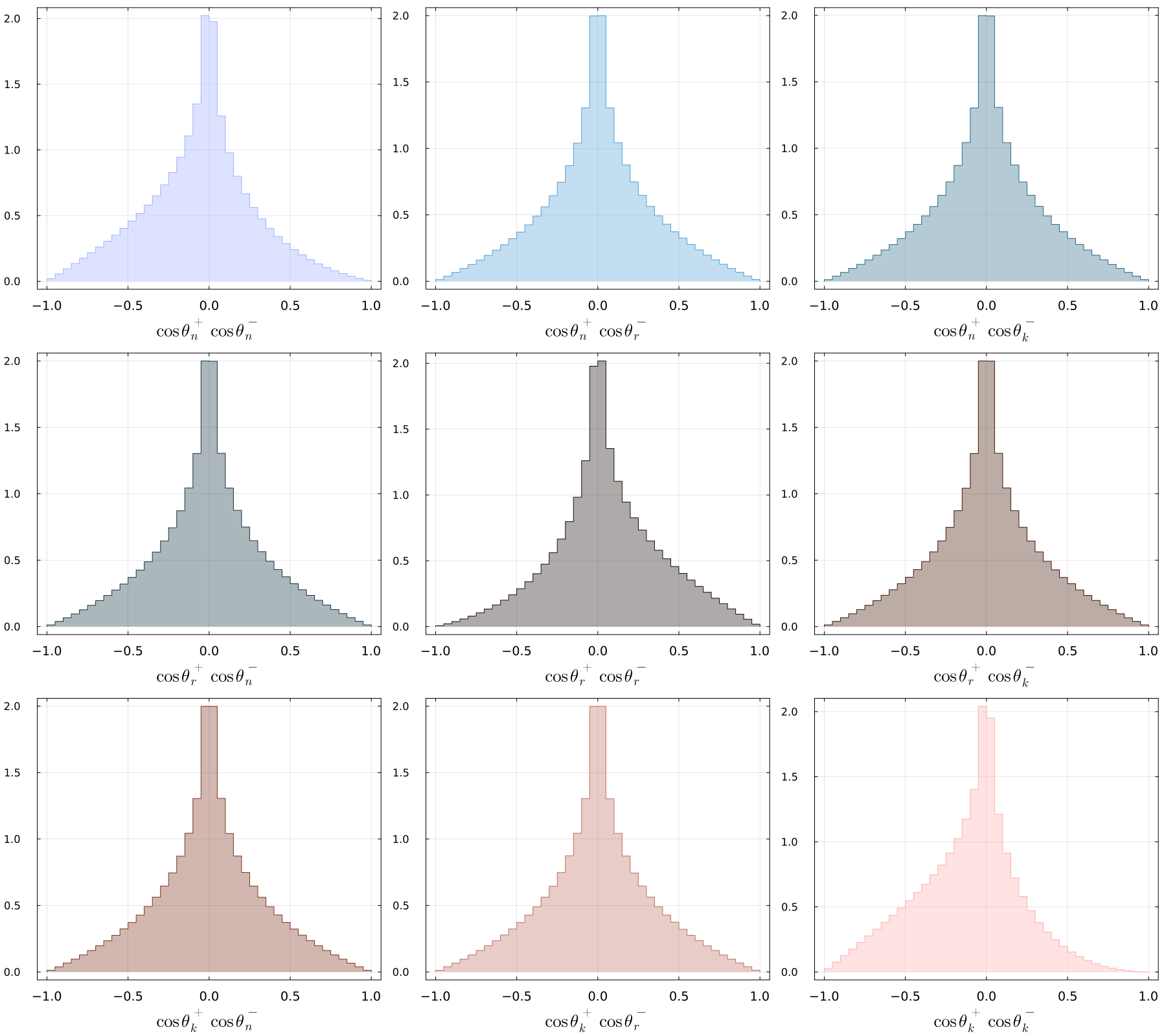}
\caption{\footnotesize Relative frequencies of the values of the \blue{products $\cos\theta^+_i \cos\theta^-_j$, $i,j=n,r,k$,} obtained with the Monte Carlo simulation of the $\tau$-lepton pairs decaying into a charged pion plus a neutrino, the momentum of which is reconstructed. The histograms are normalized to the total number of events. 
\label{fig:Cij}} 
\end{center}
\end{figure}
\subsection{Entanglement and Bell inequality violation} 

Given a two-qubit, $4\times 4$ density matrix $\rho$ as in (\ref{eq:rho}), 
its \textit{concurrence} can be explicitly constructed by using the auxiliary matrix
\begin{equation}
R=\rho \,  (\sigma_y \otimes \sigma_y) \, \rho^* \, (\sigma_y \otimes \sigma_y)\, , 
\label{auxiliary-R}
\end{equation}
where $\rho^*$ denotes the matrix with complex conjugated entries. Although non-Hermitian, the matrix $R$
has non-negative eigenvalues; denoting with $r_i$, $i=1,2,3,4$, their square roots
and assuming $r_1$ to be the largest,
the concurrence $\mathscr{C}$ can be expressed as~\cite{Wootters:PhysRevLett.80.2245}
\begin{equation}
\mathscr{C} = \max \big( 0, r_1-r_2-r_3-r_4 \big)\ .
\label{concurrence}
\end{equation}
The concurrence is a quantitative estimate of the amount of entanglement in the two-qubit system. Entanglement is present if $\mathscr{C}>0$ and maximal for $\mathscr{C}=1$.

To assess the possible violation of Bell inequalities we resort instead to the 
matrix $\CC$ in \eq{eq:xsec2d} and build, with its transpose $\CC^T$, 
the symmetric, positive, $3\times 3$ matrix
$M= \CC \CC^T$; its eigenvalues $m_1$, $m_2$, $m_3$ can be ordered in increasing order:
$m_1\geq m_2\geq m_3$. Then, the two-spin state $\rho$ in (\ref{eq:rho}) violates the Bell inequality~\cite{Bell:1964}, in  its original as well as in the equivalent Clauser-Horne-Shimony-Holt
(CHSH)~\cite{Clauser:1969ny}  form (for details see \cite{Barr:2024djo})
if and only if the sum of the two greatest eigenvalues of $M$ is strictly larger than 1, that is (\textit{Horodecki's condition}~\cite{HORODECKI1995340})
\begin{equation}
\mathfrak{m}_{12}\equiv m_1 + m_2 >1\, .
\label{eigenvalue-inequality}
\end{equation}
We take this condition as our  test of the violation of the Bell inequality. It has the advantage of automatically maximize the amount of violation without having to worry about a specific choice of basis for the polarization vectors.  

\begin{figure}[h!]
\begin{center}
\includegraphics[width=3in]{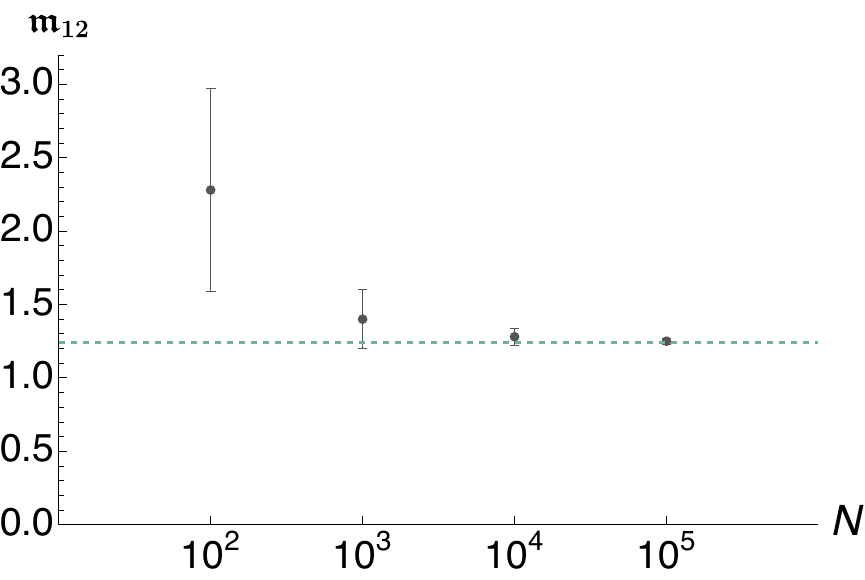} 
\caption{\footnotesize Numerical value of $\mathfrak{m}_{12}$ and its uncertainty obtained for the spin correlations of the $\tau$-lepton pairs as the size $N$ of the sample is varied. The dashed horizontal line corresponds to the analytic value. The bias is manifest for samples containing 100 events because $\mathfrak{m}_{12} $ must always be less than 2.
\label{fig:bias}  
}
\end{center}
\end{figure}
The Horodecki's condition in \eq{eigenvalue-inequality} may show a bias toward positive values when the eigenvalues are evaluated numerically on samples with a restricted statistics. The issue was addressed in \cite{Fabbrichesi:2021npl} by correcting for it in the analysis of the violation of Bell inequalities in top-quark  production at the LHC, a case  in which the number of events is limited and $\mathfrak{m}_{12} $ is biased. In this work, as in \cite{Ehataht:2023zzt}, such a correction is not necessary because, as shown in Fig~\ref{fig:bias}, the bias is negligible for samples containing at least $10^4$ events.

\subsection{Monte Carlo simulation}
\label{sub:MC}

The Monte Carlo simulation provides all the inputs required for the quantum tomography of the process, which we perform as follows.

We generate 10 million events, containing each a $\tau$ lepton pair decaying into two opposite charged pions plus the $\tau$ neutrinos, by means of \textsc{MadGraph5}~\cite{Alwall:2014hca} and the \textsc{TauDecay} library~\cite{Hagiwara:2012vz}. No cuts other than those in the default run card were applied. The events are generated at the tree level in the electroweak interaction and, therefore, the cross section and the remaining observables are expected to assume their tree-level values. The generated number of events provides an adequate benchmark to probe the capabilities of the FCC-ee to perform quantum information analyses. The expected uncertainty at the FCC-ee---where about $10^9$ events are to be collected in four years of operation \blue{with an integrated luminosity of 150 ab$^{-1}$}---can be obtained upon a rescaling by a factor $10\, \sqrt{50}$ of the \blue{statistical} uncertainties indicated by our benchmark.   
 
To make the simulation closer to the analysis with actual data, we replace the Monte Carlo truth $\tau$ lepton momenta with those obtained from the neutrino momenta reconstruction. We then sort the events into 50 independent samples (each corresponding to a pseudo-experiments with $2\times10^5$ events \blue{equivalent to an effective integrated luminosity of 17.6 fb$^{-1}$}) and reconstruct the $\BB^\pm_i$ and $\CC_{ij}$ coefficients for each sample. 

To mimic real data even further, we repeat the analysis by modeling the efficiency and uncertainty in the charged pion tracks and vertex determination of the detector prior to performing the neutrino momenta reconstruction. More details are given below. 

\subsubsection{Neutrino momenta reconstruction}

The eight unknown components of the neutrino momenta can be reconstructed by means of eight equations:
four from the sum of the $\tau$-lepton momenta, which is constrained to satisfy
\be 
p_{\tau^+}^\mu + p_{\tau^-}^\mu = p^\mu_{e^+e^-} \, ,
\ee
and four from the mass-shell conditions
\bea
(p_{\tau^+} - p_{\pi^+})^2 = m_\nu^2=0 \quad& \text{and}&\quad (p_{\tau^-} - p_{\pi^-})^2 = m_\nu^2=0\\
p_{\tau^+}^2=m_\tau^2 \quad& \text{and}&\quad p_{\tau^-}^2=m_\tau^2\, .
\eea
The system of equations is second order and a two-fold degeneracy arises (see the Appendix in \cite{Altakach:2022ywa}). 

As opposed to other processes, like top quark and $W$-boson decays, the reconstruction of the neutrino momenta and, therefore, of the $\tau$ momenta is almost perfect. The reason is that the $\tau$ lepton lives long enough to give a decay vertex that can be distinguished from the collision point. Consequently, the vector of closest approach, identified from the continuation of the trajectories of the pions emitted in the decay, can be measured and used as to resolve the two-fold degeneracy arising from the momenta reconstruction. Following~\cite{Kuhn:1993ra}, we then define the directions of the two charged pions as
\be
{\mathbf n_-} =
\frac{{\mathbf p}_-}{|{\mathbf p}_-|} \quad \text{and} \quad {\bf n_+} =
\frac{{\mathbf p}_+}{|{\mathbf p}_+|}\, ,
\ee
in which ${\mathbf p}_\pm$ are their momenta. The distance between the two decay vertices $ {\bf v_\pm}$ of the $\tau^\pm$ leptons is 
\be
{\bf d} = {\bf v_+} -  {\bf v_-} \, .
\ee
The vector of closest approach connecting the backward continuations  of the two charged pion tracks is then given by 
\be
{\bf d}_{min} = {\bf d} + \frac{ [ ({\bf d} \cdot  {\bf n_+} )( {\bf n_-} \cdot {\bf n_+}) - {\bf d} \cdot  {\bf n_-} ] \, {\bf n_-} +  [( {\bf d} \cdot  {\bf n_-} )( {\bf n_-} \cdot {\bf n_+} )- {\bf d} \cdot  {\bf n_+} ] \, {\bf n_+}}{1 - ({\bf n_-} \cdot {\bf n_+} )^2} \, .
\ee
The correct kinematic reconstruction of the $\tau$ momenta is then selected by computing ${\bf d}_{min}$ for the two solutions and comparing the results with the measured value.

\subsubsection{Detector response and initial state radiation}
The performance of what will be the actual detectors of FCC-ee can only be presumed from the specifications detailed in the current experimental proposal.

The pairs of $\tau$ leptons must be identified from the charged pions appearing in the final state. Even though the efficiency is high but not 100\%, this is not really a problem given the very large number of events available.

We model the detector resolution with \blue{two sets of uncertainties meant to mimic different systematic errors in the reconstructions. The single smearing of momenta and tracks simulate the detector response and limitations while the difference in the results obtained by means of the two sets is a measure of the experimental systematic uncertainty. We have chosen the following two sets}
\be
\frac{\sigma_{p_T}}{p_T} = 3 \times 10^{-5}\oplus 0.3 \times 10^{-3} \frac{p_T}{\rm GeV} \quad \text{and} \quad \sigma_{\theta,\phi}= 0.1 \times 10^{-3}  \, \text{rad}
\ee
\blue{or
\be
\frac{\sigma_{p_T}^\prime}{p_T} = 3 \times 10^{-5}\oplus 0.6 \times 10^{-3} \frac{p_T}{\rm GeV} \quad \text{and} \quad \sigma_{\theta,\phi}= 0.1 \times 10^{-3}  \, \text{rad}
\ee}%
for the tracks proper and
\be
\sigma_b = 3 \, \mu {\rm m} \oplus \frac{15 \, \mu{\rm m}}{\sin ^{2/3} \Theta} \frac{\rm GeV}{p_T}
\blue{\quad \text{or}\quad
\sigma_b^\prime = 5 \, \mu {\rm m} \oplus \frac{15 \, \mu{\rm m}}{\sin ^{2/3} \Theta} \frac{\rm GeV}{p_T}}
\ee
for the impact parameters and for the reconstruction of the vector of closest approach. \blue{These sets of values are taken from the envisaged IDEA tracking detector~\cite{FCC:2018evy,Azzi:2021ylt}. The specific choice we made provides a useful  benchmark to guide our analysys.} A constant smearing originates in imperfections in the detector or anomalies in signal collection, whereas the term scaling with the momentum is a noise that comes from the imperfection of the readouts. In the simulation we retain the leading contributions in the quoted uncertainties and account for the related detector effects by performing a Gaussian smearing of the Monte Carlo truth pion momenta and closest approach vector taking the nominal resolutions above as standard deviations. 

Notice that the uncertainties are small if typical pion momenta of the order of 10 GeV and $\tau$-lepton times of flight of the order of 0.1 mm are considered. Furthermore, the decay of the $\tau$ pairs will generally occur inside the beam pipe (of order 1 cm), thereby ensuring  the absence of further interactions with the detector material that could quench the spin correlation under study. 

\begin{figure}[h!]
\begin{center}
\includegraphics[width=3.5in]{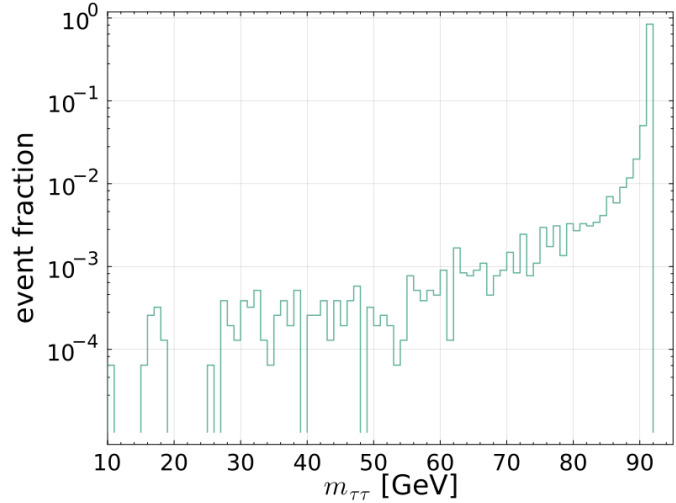}
\caption{\footnotesize Distribution of the $\tau$-lepton pair events in invariant mass after including the initial state radiation of the electrons. The last four bins, from 89 to 91 GeV, contain 99\% of the events. The count in each bin is normalized to the total number of events.
\label{fig:isr} 
}
\end{center}
\end{figure}

We also include in our Monte Carlo simulation the effect of 
Initial state radiation (ISR), which shifts the beam---and the actual center of mass  (CM)---energy as shown in Fig.~\ref{fig:isr}. The plot is obtained by using \textsc{Pythia 8}~\cite{Bierlich:2022pfr} and following the indicated statistics we pollute our dataset with events characterized by a lower CM energy up to $\sqrt{s} = 89$ GeV, corresponding to the last three bins of Fig.~\ref{fig:isr} which comprise 99\% of the events.  

The angular distributions of the pion momenta in the rest frame of the $\tau$-lepton pairs give, as the desired, the coefficients of the polarization density matrix. In the following we compare the results obtained from analytical computations to the values indicated by the Monte Carlo simulations performed with and without the ISR and detector  resolution effects. 
 
\section{Results}

{\versal Quantum tomography} gives us the means  to study entanglement and Bell inequality violation in  the process 
\be
e^{+}e^{-} \to Z, \gamma \to \tau^{+}\tau^{-} \to \pi ^{+}\pi^{-} \nu_\tau\,\bar \nu_{\tau}\, 
\ee
as it will be available at the FCC-ee.

\subsection{Theoretical quantum tomography: analytic results}

Let us first look at the analytic results for the parton level process $e^{+}e^{-} \to  \tau^{+}\tau^{-}$. This study already provides most of the information about the actual physical process, that is, the one inclusive of the hadronic decays of the $\tau$ leptons.

\begin{figure}[h!]
\begin{center}
\includegraphics[width=3in]{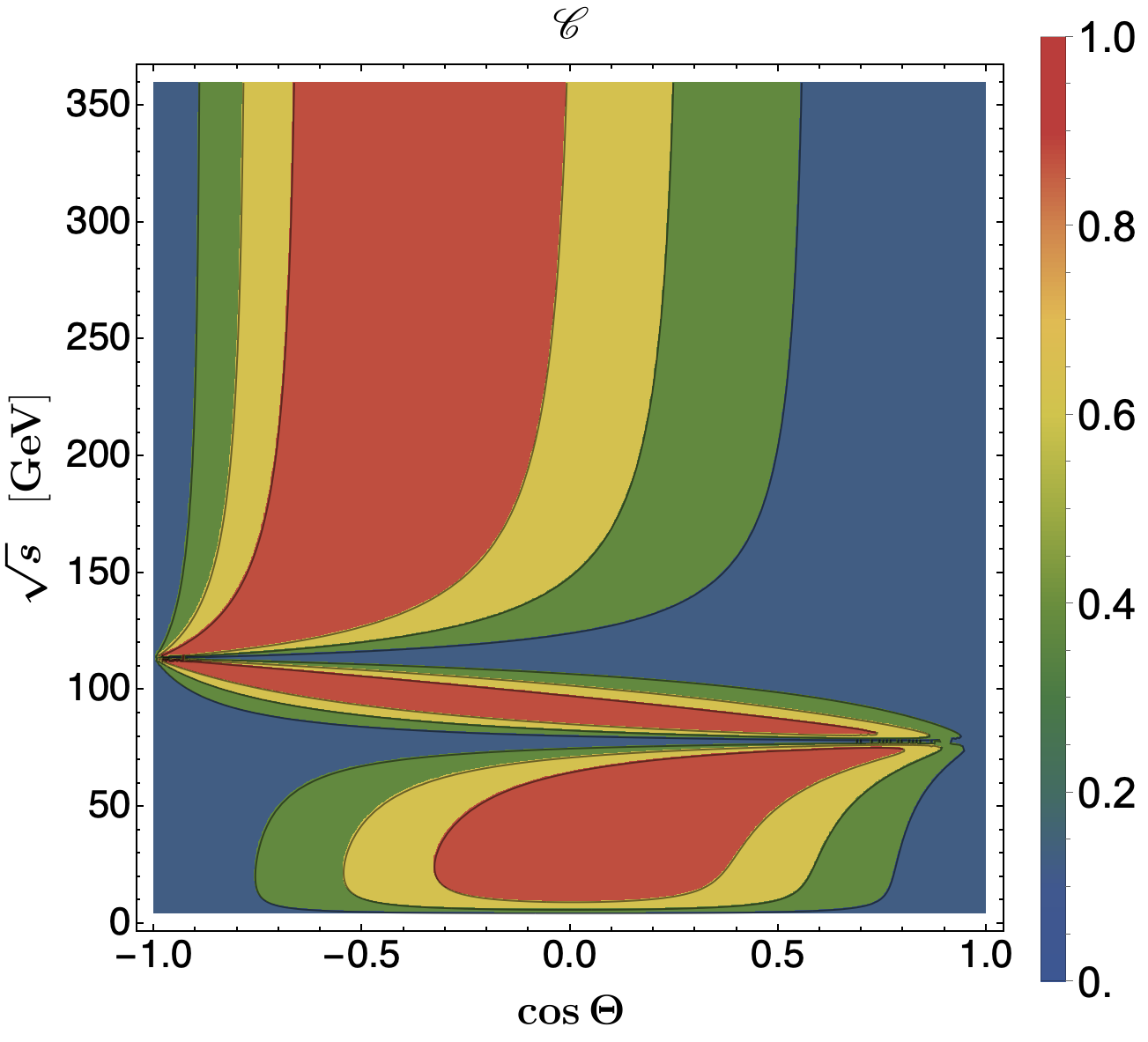}
\hspace{.5cm}
\includegraphics[width=3in]{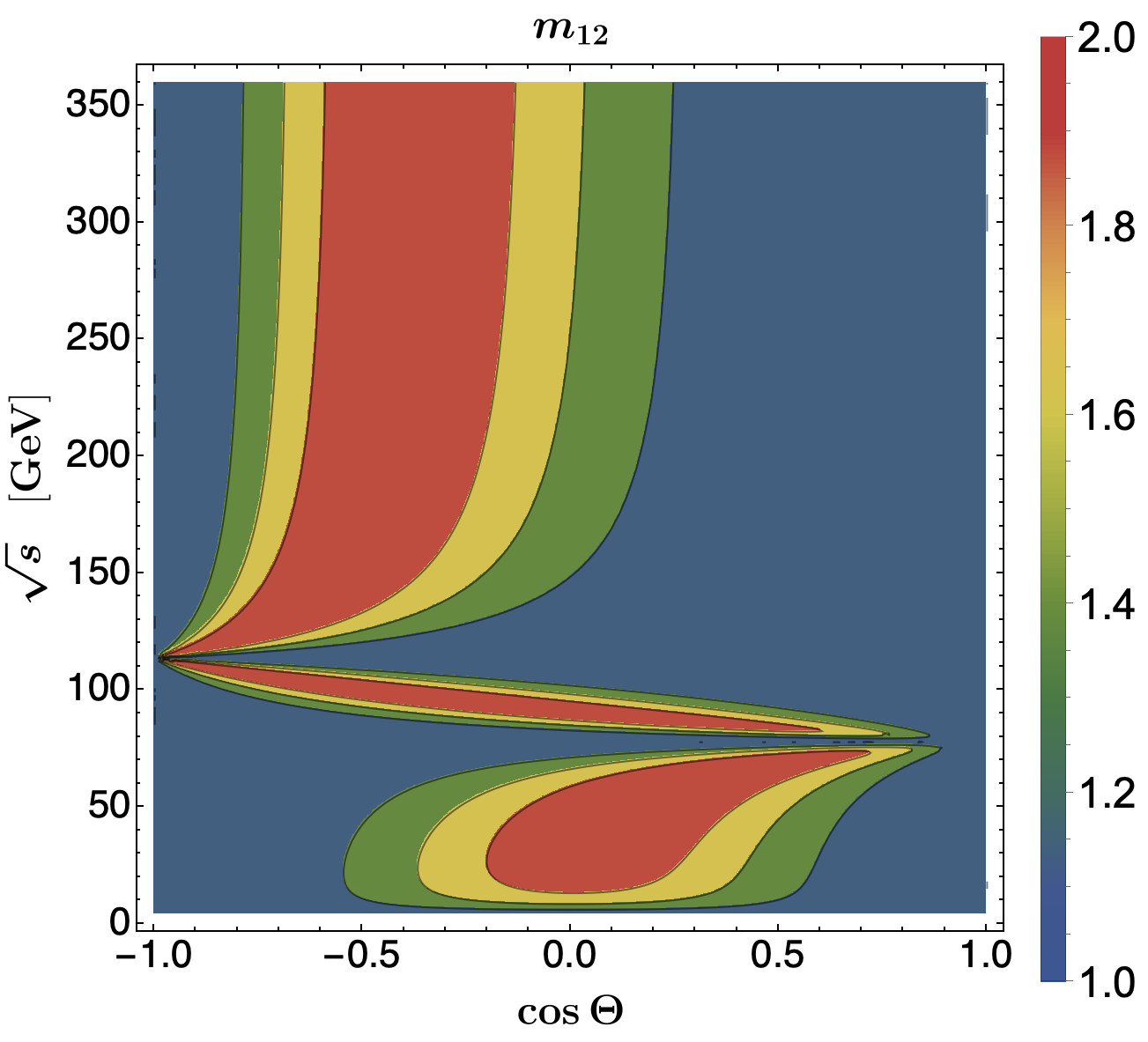}
\caption{\footnotesize Analytic solutions for the concurrence (left) and the Bell inequality violation (right) in the parton level process $e^+ e^- \to \tau^+ \tau^-$. The presence of entanglement is signaled by $\mathcal{C}>0$ and the violation of Bell inequalities by $\mathfrak{m}_{12}>1$.
\label{fig:ent}  
} 
\end{center}
\end{figure}

\begin{figure}[h!]
\begin{center}
\includegraphics[width=3.5in]{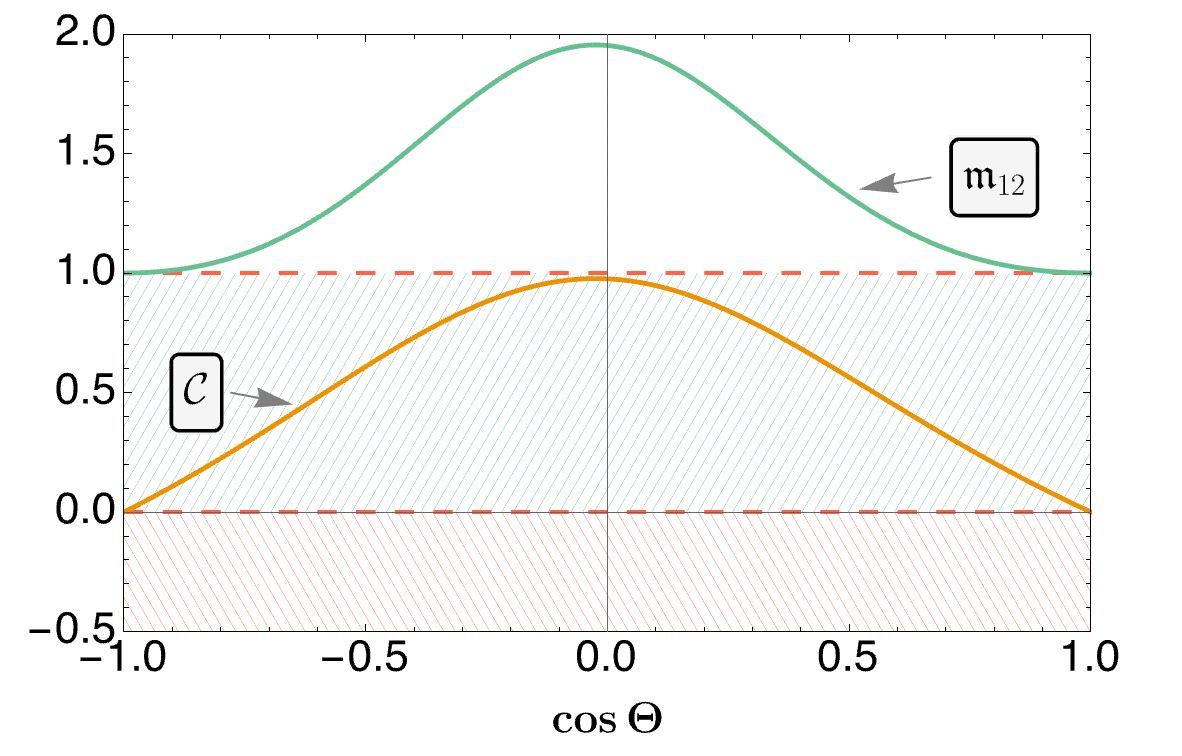}
\caption{\footnotesize Values of the concurrence $ \mathscr{C}$ and  
the Horodecki's condition $\mathfrak{m}_{12}$ as functions of the scattering angle in  the parton level process $e^+ e^- \to \tau^+ \tau^-$ for $\sqrt{s}=91.19$ GeV. The presence of entanglement is signaled by $\mathcal{C}>0$ and the violation of Bell inequalities by $\mathfrak{m}_{12}>1$.
\label{fig:ent91} 
}
\end{center}
\end{figure}

The values of the concurrence $\mathscr{C}$ and the Horodecki's condition $\mathfrak{m}_{12}$ for energies that range from the production threshold up to 350 GeV are shown in Fig.~\ref{fig:ent}. The plot shows the presence of three distinct regimes as we vary the CM energy. At low energy the process is dominated by the photon exchange, yielding the maximal value of the concurrence and of the Bell inequality violation for $\Theta = \pi/2$. At energies close to the $Z$ boson resonance the process is instead dominated by this particle. In Fig.~\ref{fig:ent} this region corresponds to the thin slice centered around $\sqrt s \simeq 90$ GeV. At higher energies both the particles equally contribute to the process and the maximal concurrence and Bell inequality violation are achieved for a different value of the scattering angle due to the different interplay among the couplings. In between different regions, interference effects reduce the entanglement.   

The result for $\sqrt{s}=10$ GeV have been already discussed in \cite{Ehataht:2023zzt} and applied to superKEKB and the physics at Belle II. Here we study the region at the $Z$-boson resonance. 

In Fig.~\ref{fig:ent91} the CM energy is set at $\sqrt{s}=91.19$ GeV and the concurrence and the Horodecki's condition are shown as functions of the scattering angle $\Theta$, formed by the $e^+$ and $\tau^+$ directions.

As for the polarization of the $\tau$ pair, we find $\BB_i^+=\BB_i^-= \BB_i$ for  $i=r, \,n, \,k$. Contrary to most of the other process that have been utilized in studies of entanglement, we have non vanishing $\BB_i$ coefficients at the leading order even though the initial beam is made of unpolarized electrons and positrons. This feature is due to the parity violating electroweak interactions and  makes the study of the $\tau$ leptons at this energy particularly interesting.
The two plots in Fig.~\ref{fig:bi_analytic} show the behavior of the coefficients $\BB_i^{\pm}$ as functions of the scattering angle $\Theta$ at $\sqrt{s} = 91.19$ GeV, as well as in a broader energy range. The characteristic pattern in the CM energy dependence is due to the interference between the photon and $Z$ boson contributions. The angular dependence is brought about by the interplay between the axial and vector couplings and the scattering angle. 

\begin{figure}[h!]
\begin{center}
\includegraphics[width=2.5in]{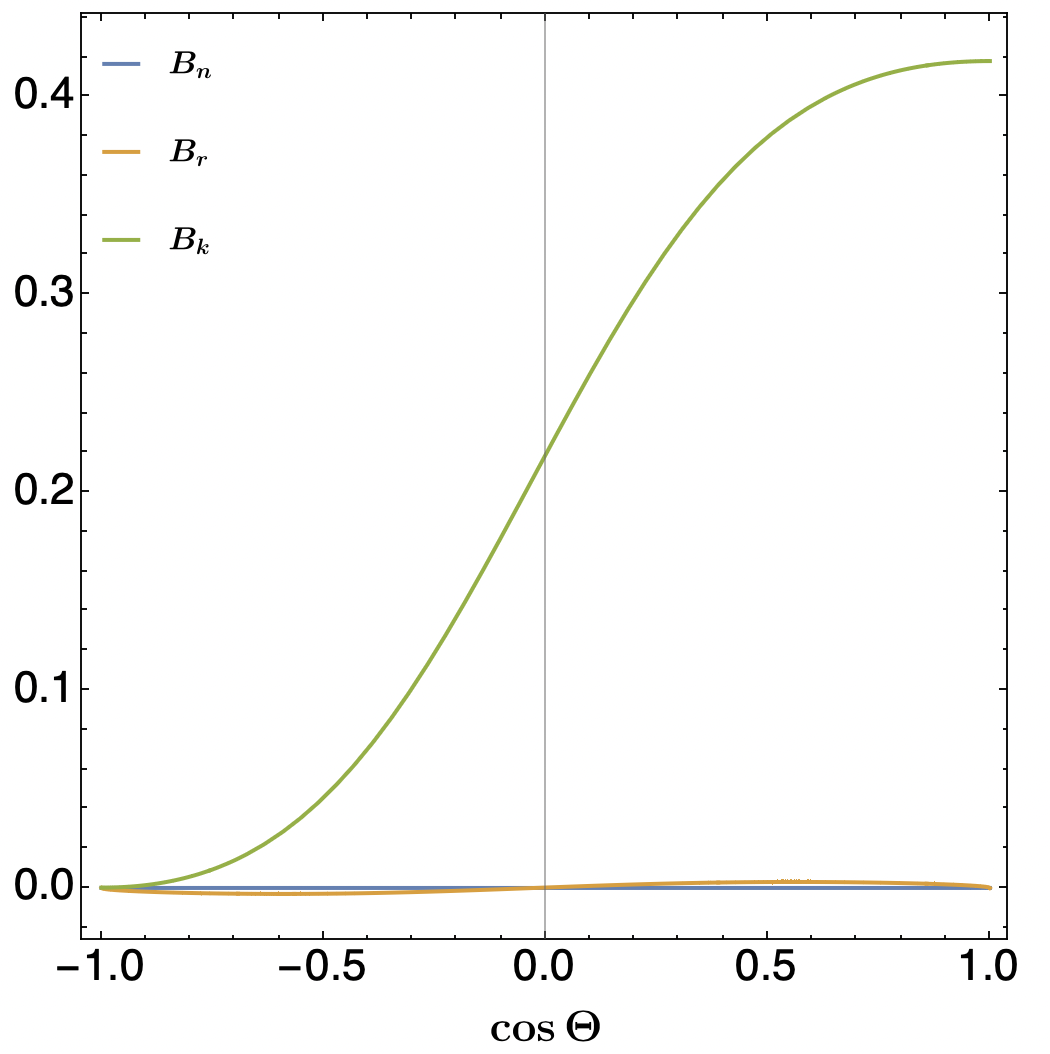}
\includegraphics[width=2.8in]{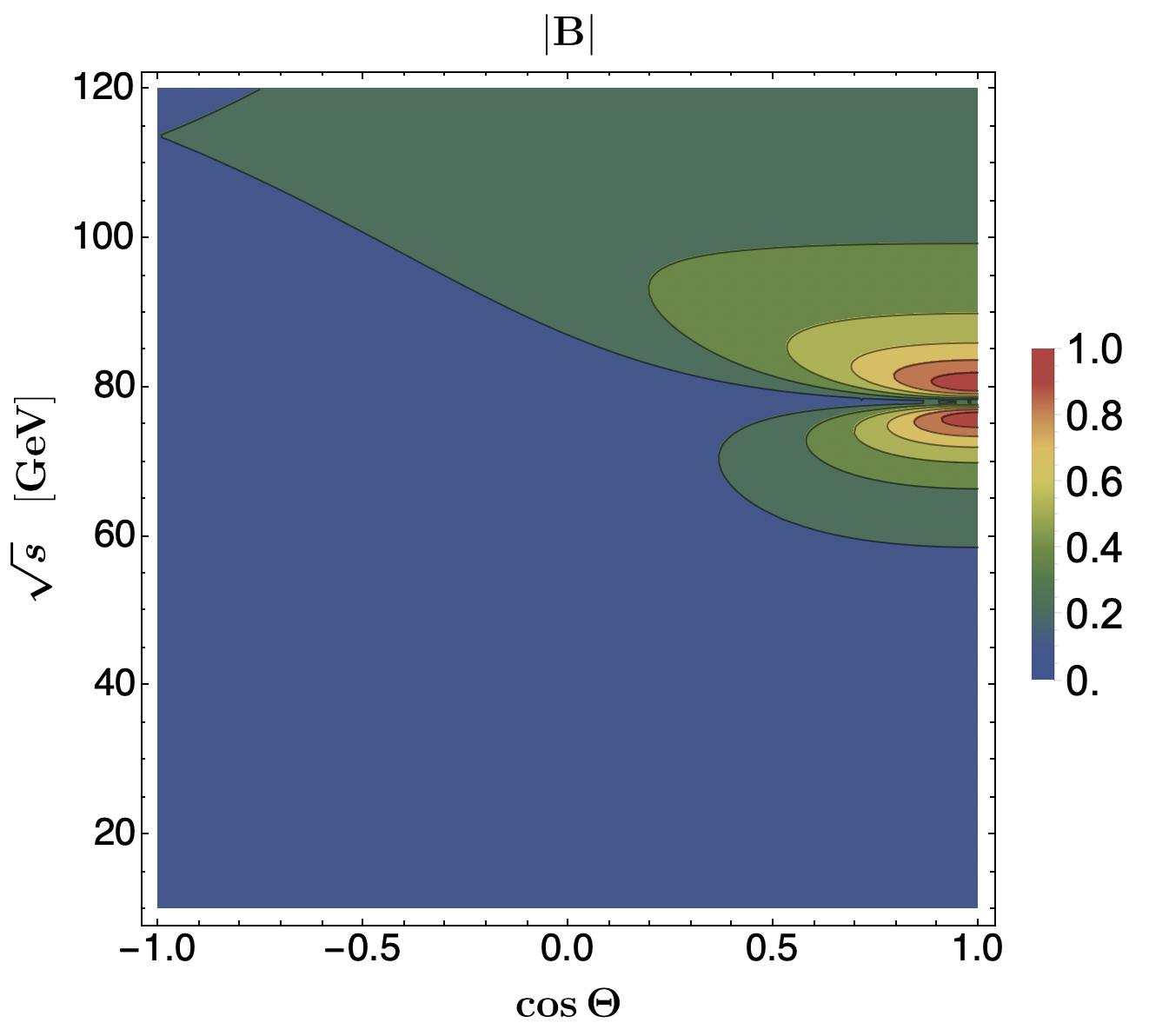}
\caption{\footnotesize Coefficients $\BB_{i}=\BB_i^+=\BB_i^-$ as functions of the scattering angle  in the parton level process $e^+ e^- \to \tau^+ \tau^-$ at $\sqrt{s} = 91.19$ GeV (left) and $|\BB|=\sqrt{\BB_k^2+\BB_n^2+\BB_n^2}$ across a broader kinematic space (right).
\label{fig:bi_analytic} 
}
\end{center}
\end{figure}

The coefficients $\CC_{ij}$ and $\BB_i$ averaged over the angular distribution of the $\tau$ pair are found to be
\be
\CC= \begin{pmatrix} 
  0.4878 &\phantom{-}  0& 0&  \\
  0 &  -0.4878 &0.0011  \\
  0 &\phantom{-} 0.0011 & 1
\end{pmatrix} \quad \quad
\BB^+ =\BB^-=\begin{pmatrix}
0\\
0.0001\\
0.2194
\end{pmatrix}\, .
\label{CandB_ana}
\ee
These values can be directly compared with the results of the Monte Carlo simulations. We find that the system is entangled and that the weighted angular averaged value of the concurrence is
\be
\mathscr{C} = 0.4878\, .
\ee
As shown in Fig~\ref{fig:ent91} the entanglement depends on the kinematic region and is maximal at $\Theta=\pi/2$. 

Similarly, the violation of the Bell inequality is given by
\be
\mathfrak{m}_{12} = 1.238\, ,
\ee
which again, is the result for the angular average, and can be enhanced by choosing a kinematic region around $\Theta=\pi/2$ (see Fig.~\ref{fig:ent91}).

While the analytic results show evidence for entanglement and Bell inequality violation, a quantitative study requires an estimate of the related uncertainties inherent to the experimental analyses. For these, we turn to the Monte Carlo simulation.

\subsection{Tomography with Monte Carlo simulation}

As explained in Section 2, we generate a total of $10^7$ events containing  $\tau$ pairs that result in two opposite charged pions and missing energy in the final states. For each of the events we reconstruct the rest frames of the $\tau^\pm$ leptons and compute the projection of their spin, indicated by the directions of the pions, on the $\{\hr, \hn, \hk\}$ basis in these frames of reference. 

The distributions in Figs.~\ref{fig:Bi} and~\ref{fig:Cij} of the coefficients obtained with the full Monte Carlo sample are not necessarily Gaussian and the skewness is what actually determines a non-vanishing average value. In order to infer a standard error and the best estimate for the value of each coefficient we then use the central limit theorem, which ensures that the sample means are normally distributed around the mean of the full Monte Carlo sample, regardless of the actual distribution of the full population. By computing the mean of the sample means and the related variance for each parameter, we are then ensured that the obtained values are faithful estimates of the corresponding true values and standard error. Having generated $10^7$  events, we opted to use 50 samples of 200 000 events each to implement the procedure \blue{(corresponding to an integrated luminosity of 17.6 fb$^{-1}$)}. We checked that the population mean of each parameter falls within the $1\sigma$ confidence interval indicated by our method. 

Omitting, for the moment, the effects of ISR and finite detector resolution,  the outlined procedure leads to the following determination of the coefficients $\CC_{ij}$  and their statistical uncertainty:
\be
\CC= \begin{pmatrix} 
 \phantom{-} 0.4810\pm 0.0061&-0.0089\pm 0.0067&-0.0001\pm0.0080\\
  -0.0074 \pm 0.0068&  -0.4806\pm 0.0065 &\phantom{-} 0.0017\pm 0.0059  \\
 \phantom{-}0.0004 \pm 0.0076& \phantom{-}0.0005\pm 0.0069& \phantom{-} 0.9982\pm 0.0073
\end{pmatrix} \,.
\ee
For the coefficients $\BB_i$ we have:
\be
\BB^+=\begin{pmatrix}
-0.0023\pm 0.0037\\
-0.0001\pm 0.0039\\
\phantom{-}0.2173\pm 0.0040
\end{pmatrix}
\quad\quad
\BB^-=\begin{pmatrix}
-0.0037\pm 0.0038\\
\phantom{-}0.0007\pm 0.0035\\
\phantom{-}0.2186\pm 0.0035
\end{pmatrix}\,.
\label{CandB_MC}
\ee

The concurrence is found to be
\be
\mathscr{C}  =  0.4815 \pm 0.0046\, , \label{C_mc}
\ee
and the Horodecki's condition for  the violation of the Bell inequality is given by
\be
\mathfrak{m}_{12} = 1.233 \pm 0.014 \, . \label{m12_mc}
\ee 

The values in \eqs{C_mc} {m12_mc} are in excellent agreement with the analytic results, proving that the size of the samples used in the analysis yields sufficient precision.

\subsubsection{Monte Carlo simulation with detector and ISR effects}

To have a better sense of the actual experimental uncertainties, we run a second Monte Carlo simulation including the smearing of the charged pion momenta and of the closest approach vector discussed in Section~\ref{sub:MC}, and contaminating the sample with events at lower $\sqrt s$ to include the effect of ISR. This procedure gives 
\be
\CC= \begin{pmatrix} 
 \phantom{-} 0.4819 \pm 0.0079 &-0.0073 \pm 0.0082 &-0.0016 \pm 0.0089\\
  -0.0066 \pm 0.0082 &  -0.4784 \pm 0.0084 & 0.0016 \pm 0.0070  \\
 - 0.0002 \pm 0.0080 & -0.0004 \pm 0.0087 &\phantom{-} 1.000 \pm 0.0074
\end{pmatrix} 
\label{C_MC_smear}
\ee

\be
\BB^+=\begin{pmatrix}
-0.0028 \pm 0.0042\\
-0.0001 \pm 0.0049\\
\phantom{-}0.2198 \pm 0.0044 \
\end{pmatrix}
\quad\quad
\BB^-=\begin{pmatrix}
-0.0039 \pm 0.0048\\
\phantom{-}0.0017 \pm 0.0049\\
\phantom{-}0.2207 \pm 0.0044
\end{pmatrix}\,.
\label{B_MC_smear}
\ee

According to the above results for the coefficients $\CC_{ij}$ and $\BB_i$, we find that the concurrence is given now by
\be
\mathscr{C} = 0.4805 \pm 0.0063  \label{C_smear}
\ee
and the Horodecki's condition by
\be
\mathfrak{m}_{12} = 1.239 \pm 0.017 \,. \label{m12_smear}
\ee


\subsubsection{Background and systematic uncertainties}

The dominant background arises from misconstruction of the $\tau$ decay channel. This is very small for the single pion channel, much smaller than for other decay channels as that into two pions. A potentially large background comes from the  presence of electron and positrons in the final state, but it can be controlled by using the impact parameter. Backgrounds arising from the process $e^+e^-\to q \bar q$ and from other sources are even smaller. We therefore feel that it is not necessary to estimate their effect for the purpose of the present analysis.

\blue{Theoretical systematic errors are either very small (those coming from NLO contributions or  the uncertainty in the value of $m_Z$) or not present (those originating from different  parton distribution functions which are not utilized in the first place) because of the EW nature of the process under consideration. } A realistic determination of the \blue{experimental} systematic uncertainties is not possible until the machine and the detectors are comprehensively understood. Yet we can partially estimate the size of the systematic errors affecting the concurrence and the Horodecki condition by accounting for ISR and the projected detector effects. 
To this end, we \blue{first} compare the absolute value of the difference between the central values of the concurrence obtained in the Monte Carlo simulation without, see \eq{C_mc},  and with detector effects and ISR, see \eq{C_smear}  
\be
\delta \mathscr{C}|_{\rm syst} = \left| 0.4815-0.4805 \right| = 0.0010\,,
\ee
similarly, for the Horodecki condition signaling the violation of the Bell inequality (cf.~\eqs{m12_mc}{m12_smear}) we have
\be
\delta \mathfrak{m}_{12}|_{\rm syst}  = \left| 1.233-1.239 \right| = 0.006\,.
\ee
\blue{These differences account for the systematic  uncertainty in the beam energy and are subdominant with respect to the statistical errors in \eq{C_smear} and \eq{m12_smear}. }

\blue{An additional contribution to the systematic errors can be obtained by using the two sets of uncertainties in the smearing procedure introduced in Section 2.3.2. These yield
\be
\delta \mathscr{C}|_{\rm syst} = \abs{0.4807-0.4805} = 0.0002\,,
\ee
and 
\be
\delta \mathfrak{m}_{12}|_{\rm syst} = \abs{1.232 - 1.230} = 0.002\,.
\ee
These contributions are, again, subdominant with respect to the statistical errors in \eq{C_smear} and \eq{m12_smear}. They become important as we rescale our statistical uncertainties toward the FCC target luminosity and will eventually come to dominate.} 

\blue{Including now also the systematic errors estimated above we find that
\be
\boxed{\mathscr{C} = 0.4805 \pm 0.0063|_{\rm stat} \pm 0.0012|_{\rm syst}\,,}  \label{C_smear2}
\ee
and the Horodecki's condition gives
\be
\boxed{\mathfrak{m}_{12} = 1.239 \pm 0.017|_{\rm stat} \pm 0.008|_{\rm syst} \,,} \label{m12_smear2}
\ee
where the systematic errors where added linearly. }

\blue{Equations~\eqs{C_smear2}{m12_smear2} are the main result of the present work. The overall significance of the Bell inequality violation (obtained with a benchmark luminosity of 17.6 fb$^{-1}$) is about 13 standard deviations once the errors are added in quadrature. The expected significance with 150 ab$^{-1}$ of data can be estimated by retaining only the systematic error in \eq{m12_smear2} and it  is about 30 standard deviations.} 

\blue{These significances can be increased by means of kinematic cuts that select events with scattering angles close to $\Theta=\pi/2$, for which the values of $\mathcal{C}$ and $\mathfrak{m}_{12}$ are larger (see Fig.~\ref{fig:ent91}). }

\subsection{Polarizations}

The coefficients $\BB_i^\pm$ are directly related to the polarizations of the $\tau$ leptons. Analytically, we find that these coefficients are equal, $\BB^+_i = \BB^-_i$ for  $i=r, \,n, \,k$.

From the Monte Carlo simulation with detector effects and ISR included \blue{(corresponding to data collected with a benchmark integrated luminosity of 17.6 fb$^{-1}$)}, we take the polarization to be the arithmetic average and obtain
 \be
\boxed{\langle P \rangle_\tau = \frac{1}{2} (\BB_k^+ + \BB_k^-) = 0.2203 \blue{ \pm 0.0044|_{\rm stat}\pm 0.0008|_{\rm syst}}\, ,} \label{P} 
\ee
while $\BB_n^\pm$ and $\BB_r^\pm$ are consistent with their vanishing analytic value, and the overall sign depends on the frame of reference. \blue{The systematic error is obtained as discussed in the previous section.} The polarization in \eq{P} corresponds to the tree-level value, consistently with the input value used for the Weinberg angle---see Sec.~\ref{subs:WA}---whereas the current measured value is $0.1476\pm 0.0108$~\cite{L3:1998oan}.

\subsubsection{The interplay between polarizations and entanglement}

The components $\BB_i$ of the polarization and the coefficients $\CC_{ij}$ of the spin correlations are closely related, for they both enter the density matrix which must satisfy positivity and normalization conditions. The entanglement is the largest for vanishing polarizations and vice versa. This relationship explains the maximum entanglement shown by amplitudes in parity conserving theories, such as QED~\cite{Cervera-Lierta:2017tdt}, that have vanishing polarizations. This also happens in electroweak processes if the parity violating structure of the lepton current is suppressed, as it is known to be for the special, and unphysical, choice $\sin \theta_W=0.5$~\cite{Cervera-Lierta:2017tdt}. For this value, the vector coupling of leptons vanishes and the current conserves parity. The cancellation is not possible for any common value of the Weinberg angle once quark currents are involved because of the different charges. For this reason,  the requirement of maximum entanglement (or the vanishing of the polarizations, which amounts to the same), though appealing, can hardly be promoted to a general principle from which to constrain a Standard Model parameters like the Weinberg angle itself.

\subsubsection{Weinberg angle}
\label{subs:WA}

The Weinberg angle can be  obtained from the polarization vector $P_\tau$ using the relation~\cite{LEPbook}
\be
P_\tau ( \cos \Theta) =  \frac{ {\cal A}_\tau\, \left( 1 + \cos^2 \Theta \right) + 2 \cos \Theta \,  {\cal A}_e}{1 + \cos^2 \Theta  + 
 2 \cos \Theta\,  {\cal A}_\tau {\cal A}_e} \, , \label{PA}
 \ee
 with, 
 \be
   {\cal A} = {\cal A}_e =   {\cal A}_\tau = \frac{ 2\,(1 - 4 \sin^2 \theta_W)}{1+(1 - 4 \sin^2 \theta_W)^2} \, ,
 \label{a}
\ee
assuming lepton flavor universality.

We compute the Weinberg angle by taking the weighted angular average of  \eq{PA} and finding ${\cal A}$ by solving the equation after inserting the value of $\langle P \rangle_\tau$ from \eq{P}. Inserting the value of  $ {\cal A}$ so obtained into \eq{a} yields the equation
\be
0.2190  =\frac{ 2\,(1 - 4 \sin^2 \theta_W)}{1+(1 - 4 \sin^2 \theta_W)^2} \, , \label{w}
\ee
 which shows that the uncertainty in the coefficient $\BB_i$ is reduced by a factor of about 8 in the error propagation.
 The central value for the Weinberg angle that solves \eq{w} corresponds to the tree level value:
 \be
\boxed{ \sin^2\theta_W = 0.2223 \blue{\pm 0.0006|_{\rm stat}\pm 0.0001|_{\rm syst}}\, ,}
 \ee
consistently with the inputs of our Monte Carlo simulation.

\blue{Rescaling the statistical uncertainty of our sample (obtained with a benchmark luminosity of 17.6 fb$^{-1}$) to account for the expected FCC-ee yield, we obtain an uncertainty of $1\times 10^{-4}$, completely dominated by the systematics.} This uncertainty is \blue{of the same order of magnitude of the current value~\cite{Workman:2022ynf}. The report~\cite{FCC:2018evy} cites uncertainties in  determination of the Weinberg angle between $10^{-5}$ and $10^{-6}$ as one of the goals for the FCC-ee machine---such a value can only be achieved if the systematic error indicated by our estimate is reduced. }
 
\subsection{Anomalous couplings}

The $\tau$ lepton---the heaviest among the leptons, as the top quark among the quarks---could be the likeliest to show a behavior departing from that described by the Standard Model.
The most general electroweak  Lorentz-invariant vertex $\Gamma^{\mu}$ between the $Z$-boson and  the $\tau$ lepton up to dimension five operators can be written as
\begin{multline}
i\,\frac{g}{2\,\cos \theta_W} \,\bar \tau \,\Gamma^\mu(q^2) \,\tau \,Z_\mu(q)
=\\
i \,\,\frac{g}{2\,\cos \theta_W} \, \bar \tau \left[ \gamma^\mu F_1^V(q^2)  +\gamma^\mu \gamma_5 F_1^A(q^2) +\frac{i \sigma^{\mu\nu}q_\nu}{2 m_\tau} F_2(q^2) 
+  \frac{\sigma^{\mu\nu} \gamma_5 q_\nu}{2 m_\tau} F_3(q^2)  \right]   \tau\, Z_\mu(q) \, , \label{int}
\end{multline}
with $F_1^V(0) = g_V=-1/2 + 2\, \sin^2\theta_W$ and $F_1^A(0) = -g_A=1/2$. $F_2(0)$ is the magnetic  and $F_3(0)$ the electric dipole moment. We parametrize the first two form factors in terms of a Taylor expansion as
\be
F_1^{V,A} (q^2) = F_1^{V,A}(0) + \frac{q^2}{m_Z^2} C_1^{V,A}\, ,
\ee
and give limits on the coefficients $ C_1^{V,A}$ at $q^2=m_Z^2$.  These form factors act as anomalous couplings of the $\tau$ lepton to the $Z$ boson.

\subsubsection{Observables}

 Our strategy to constrain the form factors in \eq{int} exploits the polarization density matrix, which can be experimentally reconstructed through quantum tomography and gives a bird's-eye view of the possible observables available for a given process. 
 
 The method has been previously used to constrain physics beyond the SM affecting the top-quark~\cite{Aoude:2022imd,Fabbrichesi:2022ovb} and $\tau$ pair~\cite{Fabbrichesi:2022ovb} production at the LHC and at superKEKB~\cite{Fabbrichesi:2024xtq}, or yielding Higgs anomalous couplings to $\tau$ leptons~\cite{Altakach:2022ywa} and gauge bosons~\cite{Fabbrichesi:2023jep,Bernal:2023ruk,Aoude:2023hxv}.

For the present case, we use the entries of the polarization density matrix to define three observables used to constrain the parameters in \eq{int}: one observable is the concurrence $\mathscr{C}$ defined in \eq{concurrence} and it measures the entanglement in the spin states of the produced $\tau$ pairs, another--- specific to the CP-violating contributes---is related to triple products involving one momentum and the spin vectors of the $\tau$ leptons:
\be
\mathscr{C}_{odd}=\sum_{\substack{i< j}} \Big| \CC_{ij} -\CC_{ji} \Big| \, , \label{CPodd} 
\ee
and the third is the total cross section: 
\be
\sigma_T = \frac{1}{64 \pi^{2}\, s}\int \di \Omega \frac{|{\cal M} |^{2}}{4} \sqrt{1 - \frac{4 m_\tau^2}{s}}\,,
\ee
in which we neglect the electron mass and take $\sqrt{s}=91.19$ GeV.

\subsubsection{Limits}

\begin{figure}[h!]
\begin{center}
\includegraphics[width=3in]{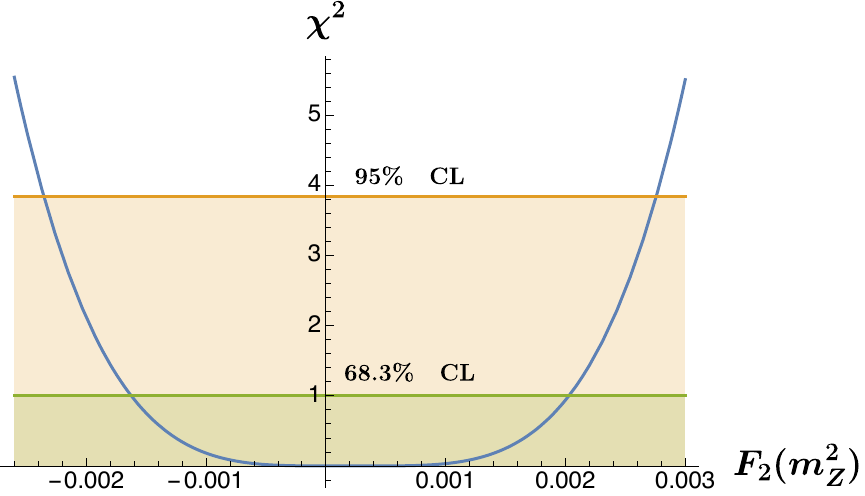}
\includegraphics[width=3in]{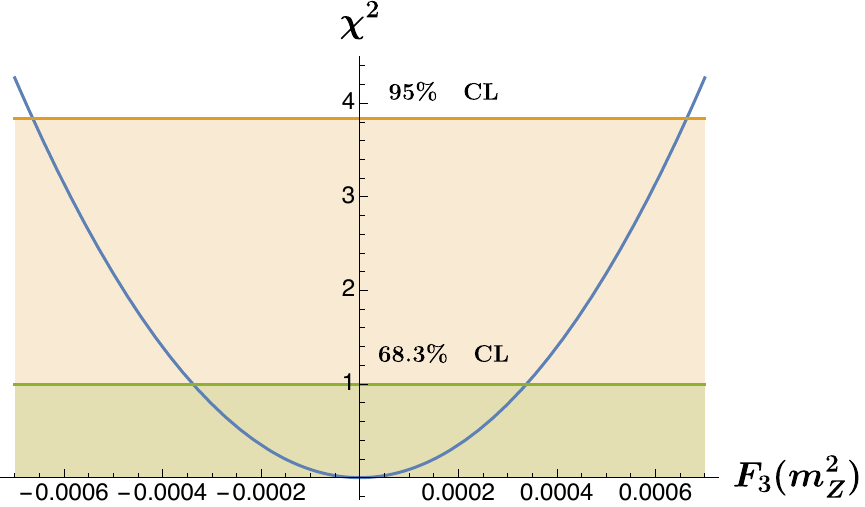}
\caption{\footnotesize $\chi^2$ test for the form factor $F_2(m_Z^2)$ and $F_3^V(m_Z^2)$. The limits for $F_2(m_Z^2)$ are obtained by means of the concurrence, those for  the form factor $F_3^V(m_Z^2)$ by means of the operator $\mathscr{C}_{odd}$.
\label{fig:limits1} 
}
\end{center}
\end{figure}

\begin{figure}[h!]
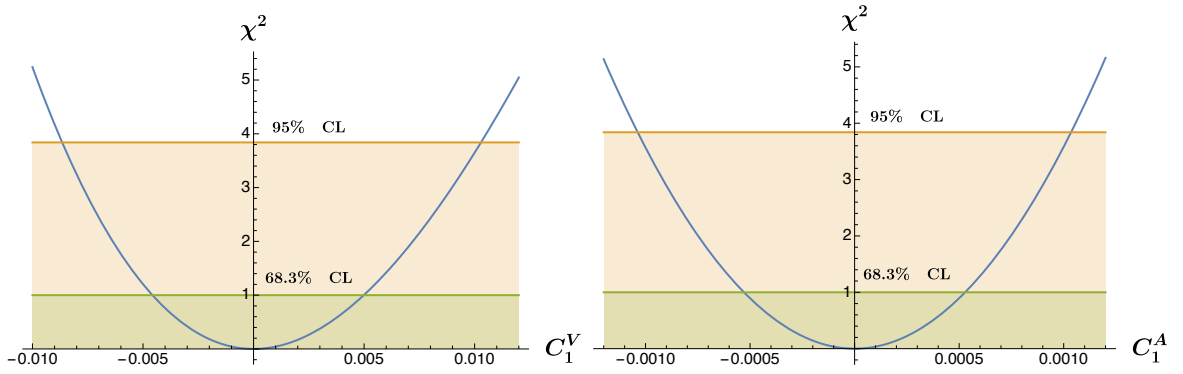

\begin{center}
\includegraphics[width=3in]{./f1V}
\includegraphics[width=3in]{./f1A}
\caption{\footnotesize  $\chi^2$ test for the form factor $C_1^V$ (left) and $C_1^A$ (right). Both limits are obtained by means of the cross section.
\label{fig:limits2} 
}
\end{center}
\end{figure}

For a sample of 200.000 events, the expected value and uncertainty on the  cross section is $(22.83 \pm 0.05)$ pb. For the same sample, from Section 3.2.1, we find an uncertainty on $\CC_{ij}$, and therefore on $\mathscr{C}_{odd}$, of at most 0.009, and on the concurrence $\mathscr{C}$ of  0.006. We use these uncertainties in the determination of the limits on the form factors. \blue{For the  number of events actually expected at the FCC-ee, we rescale the statistical errors by $10 \sqrt{50}$, which therefore becomes negligible, and use the systematic errors as the whole uncertainties, namely, a 1 per mille uncertainty in the cross section, $0.006$ for the operator $\mathscr{C}_{odd}$ and $0.001$ for the concurrence $\mathscr{C}$.} 

The operators introduced in the previous section, generically denoted here as  $\mathscr{O}_a [F_i(m_Z^2)]$, depend on the electroweak form factors $F_i(m_Z^2)$ and $\mathscr{O}_{SM}$ are the  values of the operators for the Standard Model. To constrain the couplings, we introduce a $\chi^2$ test set for a (68.3) 95\% C.L.
\be
\left[ \frac{\mathscr{O}_a[F_i(m_Z^2)]-\mathscr{O}_{SM}}{\sigma_{a}}\right]^2 \leq (1.00)\; 3.84 \,,  \label{chi2}
\ee
in which we set the uncertainties $\sigma_a$ for the operators $\mathscr{O}_a$ at the values discussed in the previous paragraph.

The bounds on each coupling can be extracted from Fig.~\ref{fig:limits1}  and \ref{fig:limits2} \blue{ in the case of 200 000 events and from similar figures in the case of $10^9$ events. The corresponding} values obtained from the 95\% confidence interval are reported in Table~\ref{tab:couplings}. Though the quoted bounds neglect possible correlations among the parameters, they are sufficient to support our claim that quantum information observables can be used to effectively constrain new physics effects.
   \begin{table}[t]
  \tablestyle[sansboldbw]
  \begin{tabular}{cccccc}
  \theadstart
  \hspace{0.2cm} \thead   $\mathscr{O}_a$ \hspace{0.2cm} & \thead  \hspace{0.5cm} $\sigma_a^{I}$  \hspace{0.2cm} &      \thead  \hspace{0.2cm}  limits I (L = 17.6 fb$^{-1}$)  \hspace{0.2cm} &   \thead  \hspace{0.5cm} $\sigma_a^{II}$  \hspace{0.2cm} &      \thead  \hspace{0.2cm}  limits II (L = 150 ab$^{-1}$)  \hspace{0.2cm} \tabularnewline
  \tbody
$\mathscr{C}$ & 0.006  &$  -0.002 \leq F_2(m_Z^2) \leq  0.003 $ & 0.001 &  $-0.001 \leq F_2(m_Z^2) \leq  0.001$\\
$\mathscr{C}_{odd}$  & 0.009 & $ -0.001 \leq F_3(m_Z^2) \leq 0.001$ & 0.006& $ -0.0004 \leq F_3(m_Z^2) \leq 0.0005$ \\
 $\sigma_T$ & 0.05 pb &$ -0.009 \leq C_1^V \leq 0.010$  & 0.02 pb &$ -0.004\leq C_1^V \leq 0.004$  \\
$\sigma_T$ & 0.05 pb  &$-0.001 \leq C_1^A \leq 0.001 $ &0.02 pb& $-0.0004 \leq C_1^A \leq 0.0004 $\\
  \tend
  \end{tabular}
  \caption{\footnotesize \label{tab:couplings} \textrm{Bounds obtained at  the 95\%  confidence level for the form factors, as  shown in Figures~\ref{fig:limits1} and \ref{fig:limits2}, neglecting correlations. \blue{The values I refer to a sample of 200 000 events, corresponding to a luminosity $L$ of 17.6 fb$^{-1}$; the values II are for the sample of $10^{9}$ events expected at the FCC after 4 years of operation with an integrated luminosity of 150 ab$^{-1}$. The uncertainty in the latter case is given by the estimated systematic error (we set that for the cross section at 1 per mille.)}}}
  \end{table}
 
In Table~\ref{tab:couplings} we report the limits as obtained with the most effective operator. For the form factor $F_2(m_Z^2)$, we could have  included those obtained with the cross section instead of the concurrence.  The cross section provides a slightly stronger bound because it has a smaller relative uncertainty. This does not mean that the cross section is a more sensitive observable than the concurrence. Fig.~\ref{fig:comp} shows that the limits from the concurrence are stronger than those from the cross section when a common relative uncertainty is considered.
 
\begin{figure}[h!]
\begin{center}
\includegraphics[width=3.5in]{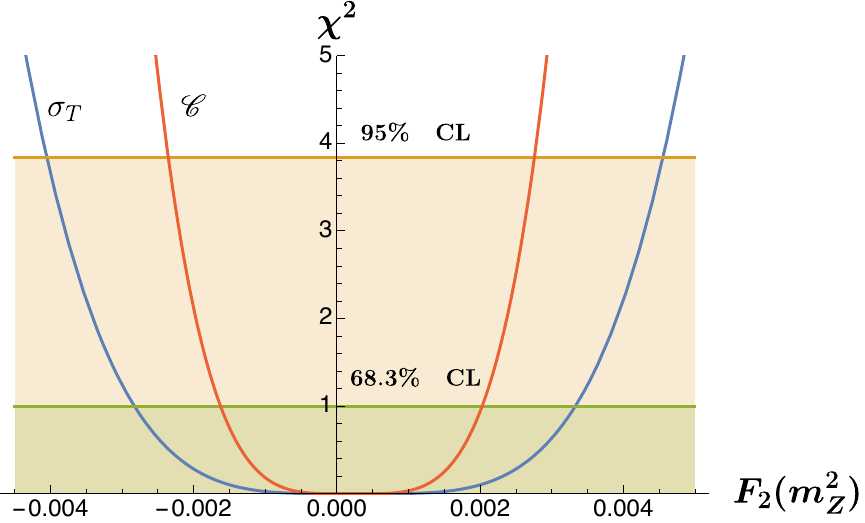}
\caption{\footnotesize  The $\chi^2$ tests for the form factor $F_2^V(m_Z^2)$ using the cross section ($\sigma_T$) and  the concurrence ($\mathscr{C}$) show that the latter provides more stringent limits when equal relative uncertainties are considered.
\label{fig:comp} 
}
\end{center}
\end{figure}

The limit on $F_3(m_Z^2)$ in Table~\ref{tab:couplings} corresponds to a limit on the CP violating weak dipole moment of $5.6 \times 10^{-18}$ e cm for the $\tau$ lepton. The corresponding limit from LEP is equal to  $1.4 \times 10^{-17}$~\cite{OPAL:1992pcg,ALEPH:1992jnn}. These limits are comparable since they are obtained with a similar number of events and utilize similar observables built out of a CP-odd triple product. Rescaling our limits to the FCC-ee luminosity, we find $|F_3(m_Z^2)| \leq \blue{1.1 \times 10^{-19}}$ e cm, which is one order of magnitude better than the current limit: $3.6 \times 10^{-18}$ e cm~\cite{Workman:2022ynf}.
 
Table~\ref{tab:couplings} shows the bounds obtained with pseudo-experiments containing each 200 000 events \blue{(second column) and} the corresponding projections base on the full dataset of $10^9$ expected at the FCC-ee  \blue{(fourth column), in which only the systematic uncertainty is considered since the  statistical errors are negligible}  upon rescaling by a factor $10 \sqrt{50}$.
   
\vskip2cm
\section{Outlook}

{\versal  The FCC-ee can}, after 4 years of operation at the energy of the $Z$-boson mass, test quantum entanglement and Bell inequality violation in $\tau$-lepton pairs with extraordinary precision. Already with the reduced sample we have used in our analysis the significance is well in excess of five standard deviations.  \blue{The same large  significance is obtained with the full  dataset of $10^{9}$ events---at which point the systematic uncertainty will dominate and any further progress will depend on better understanding this uncertainty. The same holds for the determination by quantum tomography  of the Weinberg angle from the $\tau$-lepton polarizations and constraining the form factors entering the electroweak neutral current of the $\tau$ leptons.} We believe that the results we have presented make a convincing case for including the quantum tomography of the $\tau$ leptons in the physics goals of the FCC-ee.  

\section*{Acknowledgements}
{\small
The authors thank Michele Pinamonti and Christian Veelken for discussions. L.M. is supported by the Estonian Research Council grants PRG803, RVTT3 and by the CoE program grant TK202 ``Fundamental Universe'’.}

\newpage

\begin{multicols}{2}
\small
\bibliographystyle{JHEP}   
\bibliography{taus.bib} 

\end{multicols}
\end{document}